\documentclass[12pt]{article} 

\usepackage[utf8]{inputenc}
\usepackage[T1]{fontenc}
\usepackage{geometry}
\usepackage{mathptmx} % Times New Roman font 
\usepackage{graphicx}
\usepackage{amsmath}
\usepackage{hyperref}
\usepackage{longtable}
\usepackage[normalem]{ulem}
 
\usepackage[final]{changes}

\title{T-cell receptor binding prediction: A machine learning revolution}

\author{Anna Weber\footnote{IBM Research Europe, 8803 Rüschlikon, Switzerland.} \and Aurélien Pélissier\footnote{Institute of Computational Life Sciences, Zürich University of Applied Sciences (ZHAW), 8820 Wädenswil, Switzerland.} \and Mar\'ia Rodr\'iguez Mart\'inez\footnote{Biomedical Informatics and Data Science, Yale School of Medicine, New Haven, CT 06510, USA.} \,
\footnote{Correspondence to maria.rodriguezmartinez@yale.edu.} }

\date{\today}  

\begin{document}

\maketitle 

\section*{Abstract}

Recent advancements in immune sequencing and experimental techniques are generating extensive T cell receptor (TCR) repertoire data, enabling the development of models to predict TCR binding specificity. Despite the computational challenges posed by the vast diversity of TCRs and epitopes, significant progress has been made. This review explores the evolution of computational models designed for this task, emphasizing machine learning efforts, including early unsupervised clustering approaches, supervised models, and recent applications of Protein Language Models (PLMs), deep learning models pretrained on extensive collections of unlabeled protein sequences that capture crucial biological properties.

We survey the most prominent models in each category and offer a critical discussion on recurrent challenges, including the lack of generalization to new epitopes, dataset biases, and shortcomings in model validation designs.
Focusing on PLMs, we discuss the transformative impact of Transformer-based protein models in bioinformatics, particularly in TCR specificity analysis. We discuss recent studies that exploit PLMs to deliver notably competitive performances in TCR-related tasks, while also examining current limitations and future directions.
Lastly, we address the pressing need for improved interpretability in these often opaque models, and examine current efforts to extract biological insights from large black box models.

\vspace{1cm}

\noindent \textbf{Keyword}: Machine Learning; T~cell Receptor; Specificity Prediction; Protein Language Models; Interpretability.

\vspace{.5cm}

\section{Background}

T~cells are an essential component of the adaptive immune system due to their ability to orchestrate targeted, effective immune responses through cell-based and cytokine-release mechanisms. While T~cell functions are diverse, their activation, differentiation, proliferation, and function are all governed by their T~cell receptors (TCR), which enable them to recognize non-self antigens arising from infectious agents or diseased cells~\cite{shah_t_2021}.

To face a diverse and ever-evolving array of antigens, the immune system has evolved the capability to generate a huge array of distinct TCRs. This diversity is achieved through a random process of DNA rearrangement, which involves the recombination of the germline V, D, and J gene segments and the deletion and insertion of nucleotides at the V(D)J junctions. 
While the theoretical diversity of different TCRs is estimated to be as high as  $10^{19}$~\cite{dupic_genesis_2019}, the realized diversity in an individual is much smaller, typically ranging between $10^{6}$ and $10^{10}$~\cite{Laydon2015}.
%The entirety of all TCRs of a person is called the T~cell repertoire. 

At the molecular level, TCRs interact with peptides presented on the major histocompatibility complex (MHC), a complex commonly referred to as pMHC. 
Although the interaction between pMHC and TCR is highly specific, a single TCR can often recognize multiple pMHC complexes. Indeed, some TCRs have been shown to recognize up to a million different epitopes~\cite{Wooldridge2011}. This multivalency is necessary to ensure that the realized diversity in one individual can recognize a significantly broader array of potential antigens.

\section{T~cell receptor specificity prediction.}\label{sec:intro_tcr_specificity}

The accurate prediction of TCR-pMHC binding is crucial for accurately estimating immune responses and holds the promise to revolutionize the development of immunotherapies. For instance, the precise determination of the epitopes recognized by expanded TCR clones can aid in the identification of auto-antigens in T-cell-associated autoimmune diseases~\cite{weber_identification_nodate}, advance the development of more effective and less toxic immunotherapies, and facilitate the investigation of the pathogenic agents eliciting T-cell responses~\cite{nolan2020}. In cancer, enhancing the predictive accuracy of TCR specificity can not only improve the design of more effective T cell-based therapies~\cite{bashor_engineering_2022} but also reduce toxic side-effects caused by off-target TCR binding~\cite{linette2013}.

However, experimental methods cannot sample the vast space of potential TCRs and epitopes, and hence, significant emphasis has been placed on developing reliable computational methods to predict TCR specificity. Existing approaches can accurately classify in-distribution samples, i.e. they can predict TCR binding to epitopes already encountered by the model~\cite{meysman2023}. However, the pivotal challenge is to develop models that can generalize to novel epitopes. A major challenge stems from the scarcity of datasets containing experimentally validated TCR-epitope interactions and the low diversity of epitopes sampled.

\section{Limitations of available datasets.}\label{sec:intro_data}

TCR specificity data can be collected from various databases, such as the VDJdb~\cite{goncharov2022}, with over $\sim$70,000 TCR sequences and $\sim$1100  different epitopes as of December 2023, and McPas-TCR~\cite{tickotsky2017}, with a manually curated set of $\sim$40,000 pairs. Newer datasets are also rapidly becoming available, such as the MIRA dataset,  published during the COVID pandemic and including over 135,000 TCRs binding various COVID-19 epitopes~\cite{nolan2020}.

However, current datasets exhibit serious limitations. First, while bulk sequencing of T cells is high-throughput and cost-effective, it cannot detect paired $\alpha$ and $\beta$ chain sequences. New single-cell technologies can generate paired-chain data, yet they are costly and remain relatively underrepresented in public datasets. Currently, only a minor fraction of samples in VDJdb and none in the MIRA datasets provide paired-chain data.
Furthermore, the experimental methods predominantly rely on known target pMHC complexes, skewing the datasets towards  TCRs that recognize a limited number of epitopes, predominantly of viral origin and associated with the 3-6 most common HLA alleles. Finally, the datasets also show a significant bias in epitope diversity, with just $\sim100$ antigens accounting for 70\% of TCR-antigen pairs~\cite{hudson_can_2023}.

%\added{\sout{The lack of negative data in T~cell sequencing, which focuses primarily on pMHC-labeled cells, further challenges the development of accurate supervised machine learning models. Different approaches are typically used to artificially generate non-binding TCR-epitope pairs, ranging from shuffling TCR-epitope pairs, to using naive TCR sequences, or decoy datasets. However, each of these methods introduces its own biases, and careful consideration is needed in their application to ensure the generation of negative data that accurately reflects true non-binding interactions~\cite{moris2020current}.}}

The lack of negative data in T-cell sequencing, which focuses primarily on pMHC-labeled cells, further challenges the development of accurate supervised machine learning models~\cite{dens2023pitfalls}. Experimental methods to determine TCR–epitope pairs have high specificity but low sensitivity, resulting in a high false negative rate~\cite{wang2021direct, zhang_framework_2021}. For instance, tetramer-based approaches capture only a subset of the true TCR–epitope interactions~\cite{rius2018peptide}. Due to the scarcity of true negative pairs in TCR–epitope databases, artificial negative instances must be generated to train supervised TCR–epitope prediction models.

Two common methods for generating negative TCR-epitope pairs are shuffling known positive pairs and using TCR data from unrelated experiments. The first method randomly pairs each TCR with a different epitope from the same experiment, assuming that a TCR specific to one epitope is unlikely to bind to another. However, the limited number of known epitope-TCR pairs restricts the size of the training, test, and validation sets, thus reducing the model's predictive accuracy. The second approach pairs known epitopes with random TCRs from broad sequencing experiments~\cite{gao2023}. However, this approach introduces biases, as the negative and positive pairs come from different experiments, often conducted in different labs and involving subjects from different ethnicities. Furthermore, high-performance machine learning models have been observed to memorize in some cases CDR3 sequences, leading to over-optimistic performances~\cite{grazioli2022tcr, moris2020current}. While the optimal approach to generate negatives is still debated, careful consideration is needed to ensure the generated negative pairs accurately reflect the true non-binding distribution~\cite{dens2023pitfalls, moris2020current, hudson2023}.

\section{Evolution of TCR specificity prediction models.}

Since the first release of TCR-pMHC binding data in 2017~\cite{dash2017}, multiple studies have undertaken the challenge of modeling TCR specificity, with models broadly categorized into three groups: unsupervised clustering methods, supervised classifiers, and protein language models.
Namely, in the early years of TCR-pMHC modeling (2017-2019), when the scarcity of labeled data posed challenges for training supervised models, simple clustering algorithms demonstrated the feasibility of predicting TCR specificity from sequences. In 2020 and 2021, with the increased availability of data, there was a surge in supervised models ranging from simple classifiers to deep neural networks.
In the last couple of years, the breakthrough of Large Language Models, such as OpenAI’s generative pre-trained Transformer (GPT) models~\cite{openai_gpt-4_2023}, has facilitated the emergence of Protein Language Models (PLMs) based on similar principles~\cite{brandes2022proteinbert,rives2021biological,chowdhury2022single,madani2023large}. The number of PLMs is rapidly increasing, with some of them being specifically trained on TCR sequences~\cite{wu2021,kwee2023}. 
\begin{figure}[ht]
\centering
\includegraphics[width=\linewidth]{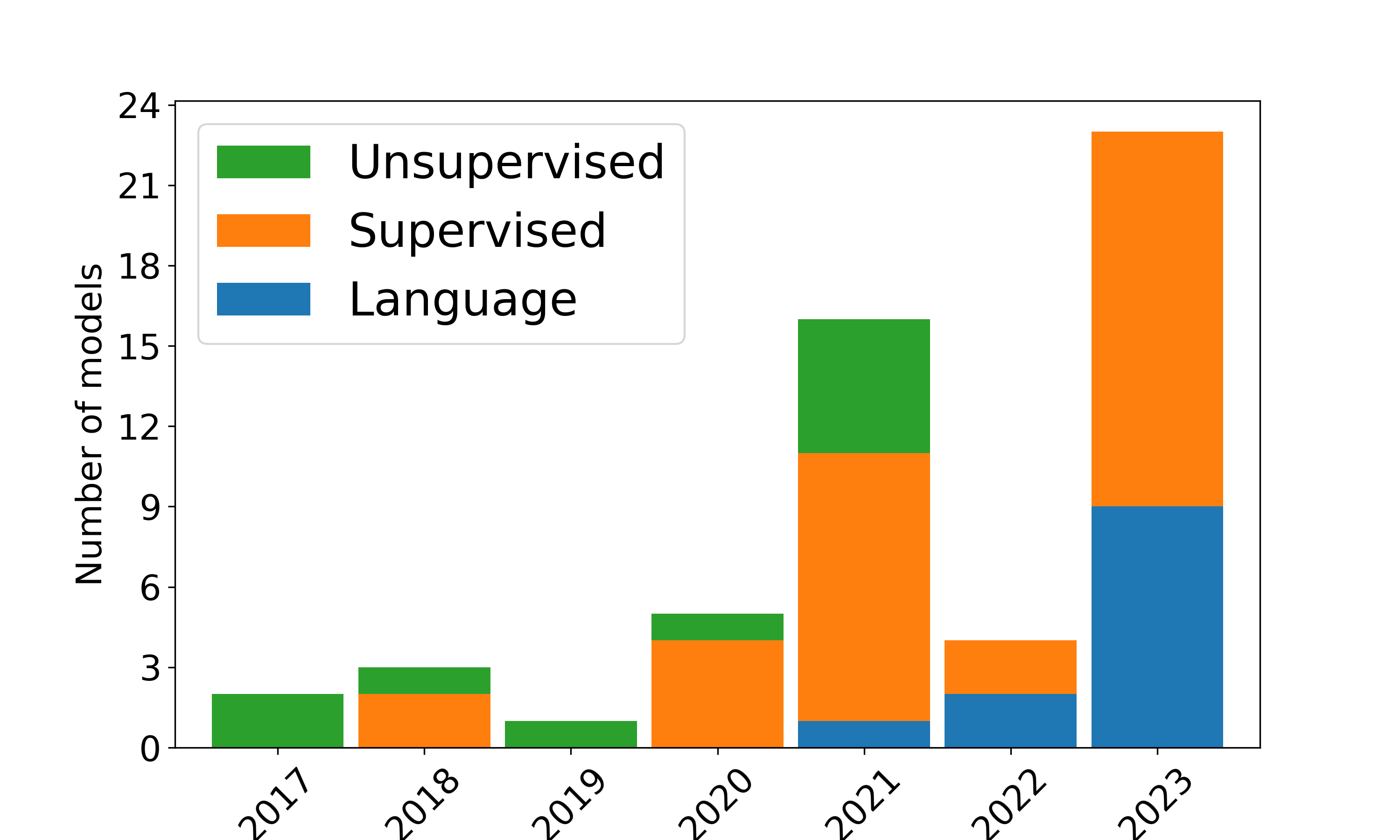}
\caption{\textbf{Publication evolution of TCR specificity models.} An overview of the models considered for this plot is shown in Table~\ref{table:models}. }
\label{fig:models_history}
\end{figure}
Fig.~\ref{fig:models_history} illustrates the evolution of modeling approaches, prompting us to refer to the different waves as generations.
In what follows, we describe these three model generations and summarize the main characteristics of prominent representatives of each.

\subsection{1st generation: unsupervised clustering}

The initial efforts in TCR specificity prediction employed unsupervised clustering methods,  under the assumption that sequence similarity, or more precisely, similarity of sequence features, correlates with specificity similarity. Under this hypothesis, clusters of TCRs with similar sequences are expected to bind to the same targets. Typically, these approaches establish a distance measure and then apply simple machine learning methods, such as K-nearest neighbors, to label test samples based on their closest training samples. 

A simple proof of concept was provided by De Neuter et al~\cite{deneuter2018} on a dataset comprising only two HIV epitopes. Namely, a random tree classifier was trained using TCR features, such as the V and J segments, CDR3 sequence length, mass, and amino acid counts, as well as biochemical features, such as CDR3 basicity, hydrophobicity, helicity, and isoelectric point. While the applications of this model were limited, it elegantly showed that specificity prediction from sequences is feasible. TCRex~\cite{gielis2019}, described in more detail in the next section,  is an advanced version of the model that uses random forest for classification. 

TCRdist, the most established TCR distance measure, was introduced by Dash et al. in 2017~\cite{dash2017}. 
TCRdist considers the pMHC contact loops of the TCR, specifically the CDR1, CDR2, and CDR3 as well as an additional variable loop labeled CDR2.5 that has been observed to make contact with pMHC in solved structures.
It then calculates the mismatch distance between the amino acid sequences of these loops using the BLOSUM62 substitution matrix and applying a predefined gap penalty. Furthermore, mismatches and gap penalties are weighted, with the CDR3 region having a higher weight to reflect its greater importance in peptide recognition. The final TCRdist score is the sum of the weighted mismatches and gap penalties, normalized by the length of the alignment.
%
%TCRdist focuses on the CDR loop regions, where the loop definitions are taken from the IMGT CDR definitions~\cite{imgt} with two key modifications: the inclusion of an additional pMHC-facing loop between CDR2 and CDR3, recognized for its interaction with pMHC in structural analyses, and the use of a trimmed version of CDR3. Additionally, the CDR3 region is assigned a higher weight of 3, underscoring its major role in the specificity and functionality of TCRs.
%
%\added{The weighted sum of these CDR loop alignment scores is then utilized as a measure of proximity, assigning TCRs the peptide specificity of their closest neighbor.}
%
Using this distance measure, it was shown that TCRs with similar specificity cluster together, and the specificity of uncharacterized TCRs can be correctly predicted based on their proximity to training sequences. A newer version incorporating several new features has been introduced as TCRdist3~\cite{mayer-blackwell2021a}. 

In the same year, Glanville and colleagues released GLIPH (Grouping of Lymphocyte Interactions by Paratope
Hotspots)~\cite{glanville2017}, a TCR clustering model based on both global similarity and local motif similarity. The model establishes an edge between TCRs that either differ by fewer than two amino acids or share a sequence motif enriched more than 10-fold relative to naive repertoires. Clusters are then determined as communities in the resulting graph. 

To enhance the clustering efficiency, ClusTCR~\cite{valkiers_clustcr_2021} leverages the Faiss Clustering Library, a library specifically developed for rapid clustering of dense vectors through efficient indexing, forming superclusters based on sequences' physicochemical properties. It also employs an optimized K-means algorithm to compute centroids and assign each sequence to its nearest centroid.
In the second phase, ClusTCR explores potential specificity groups within each supercluster. It uses sequence hashing and the Hamming distance to construct graphs of related sequences and employs the Markov clustering algorithm to detect dense network substructures, which may indicate clusters with similar specificities.

The success of these early models demonstrated that the TCR sequence encodes specificity information and led to the identification of a set of features, such as high-level sequence descriptors, edit distance, and motif sharing, which subsequent, more complex models have since utilized. Interestingly, even with the increased availability of data, distance-based approaches, such as  TCRMatch~\cite{chronister2020tcrmatch}, GIANA~\cite{zhang2021}, iSMART~\cite{zhang2020},  ELATE~\cite{dvorkin2021}, etc,  have continued to be developed and demonstrate competitive performance on many tasks. 
However, while clustering approaches are straightforward and effective in environments with limited data, they struggle to accurately represent complex nonlinear interactions and out-of-distribution data. As more data became accessible through public databases, new and more sophisticated supervised methods began to emerge.

\subsection{2nd generation: supervised models}
In the domain of supervised models, clear distinctions emerge in both the modeling approach and the formulation of the prediction task. The modeling aspect encompasses approaches ranging from nonparametric machine learning models to neural network architectures. As for the prediction task, there are two primary methods. The first treats known epitopes as distinct classes, to which TCRs are assigned. This method, used by all unsupervised clustering methods and approximately one-third of supervised models, does not utilize epitope information as input, and hence, it cannot generalize to new epitopes. Conversely, models that explicitly incorporate epitopes as input aim to predict the binding probability between any TCR and epitope based on their sequences.

Early models predominantly employed non-parametric methods and treated epitopes as class labels for TCRs. Notable examples include TCRGP~\cite{jokinen2019},  SETE (Sequence-based Ensemble learning approach for TCR Epitope binding prediction)~\cite{Tong2020}, and SwarmTCR~\cite{ehrlich_swarmtcr_2021}. TCRGP searches for similarities between TCRs using a Gaussian Process classifier with a squared exponential kernel function based on a BLOSUM encoding, i.e. an amino acid-based encoding extracted from the Blocks Substitution Matrix~\cite{henikoff_amino_1992}. SETE utilizes k-Mers of the CDR3 sequence as input features and adopts an ensemble learning approach based on decision trees to classify TCR sequences into epitope-binding classes. 

TCRex~\cite{gielis2019} uses the CDR3 amino acid sequence and V/J gene information of the TCR$\beta$ chain to predict TCR-epitope binding. TCRex employs random forest classifiers trained on 3 datasets
%, the McPAS-TCR, VDJdb, and ImmuneCODE™ databases, 
covering 100 different epitopes, including 93 viral and 7 cancer epitopes. For each epitope, TCRex trains a model using physicochemical features of the CDR3 region and one-hot encoded information about the V and J genes. It addresses the natural imbalance in epitope-specific T cells within TCR repertoires by over-sampling negative data.

Building upon TCRdist, SwarmTCR~\cite{ehrlich_swarmtcr_2021}  recognizes the variable significance of the $\alpha$ and $\beta$ chains and the various CDR regions for different peptides, and seeks to determine the optimal weights for each of the eight CDR loops, tailored to specific peptides. 
To achieve this, it utilizes Particle Swarm Optimization (PSO), an established optimization method, to compute the optimal weights for both single-cell and bulk sequencing data. These weights are then applied in a nearest-neighbor fashion similar to TCRdist to maximize the classification performance for a set of peptide labels.

A majority of supervised models, however, employ neural network architectures.
One of the first attempts, NetTCR~\cite{Jurtz2018}, used convolutional layers and separated input streams for TCR and epitope sequences, using BLOSUM encodings to vectorize amino acid sequences. Due to the low number of available training sequences at the time, the model performance was moderate, although improved versions with enhanced accuracy have been released afterwards~\cite{montemurro2021,jensen_nettcr_2023}.

TcellMatch~\cite{Fischer2020}, another early neural network model, featured a variety of layer types, including self-attention, Gated Recurrent Units, Long Short-Term Memory (LSTM), and convolutional layers. 
The authors explored the use of paired chain data versus only TCR$\beta$ data, and experimented with adding covariate data such as transcriptome and surface protein expression from single-cell experiments. Furthermore, the explicit encoding of epitopes was compared with using epitopes as class labels, with the latter method showing an improved performance.
%While they adapt embedding TCR sequences using the BLOSUM matrix, they suggest adding a learned 1x1 convolutional layer to improve the embedding. 

Published shortly afterward, ImRex~\cite{moris2020current} introduced the novel concept of representing TCRs and epitopes as visual interaction maps, facilitating the use of established computer vision techniques. The authors also investigated the model's ability to generalize predictions to new epitopes, and demonstrated that, while extrapolating to unseen epitopes remains a challenging task, ImRex was able to make predictions for epitopes that were not too dissimilar from the training data.

Both DeepTCR~\cite{sidhom2021deeptcr} and pMTnet~\cite{lu2021} exploited autoencoders to generate meaningful representations of TCR sequences in a latent space, later used as input for a classifier. This has the advantage of permitting the use of additional unlabeled data to train the encoder and decoder. DeepTCR employed variational autoencoders with convolutional layers, and used the TCR CDR3 sequence along with V and J gene information. 
The key difference in pMTnet is the use of stacked autoencoders instead of variational ones, and the explicit encoding of the pMHC molecule through a re-implemented version of the netMHCpan~\cite{nielsen_netmhcpan-30_2016}, an MHC-I binding machine-learning model.

Other natural language processing (NLP) techniques have been tested to predict TCR-peptide binding. For instance, ERGO~\cite{springer_prediction_2020} (pEptide tcR matchinG predictiOn) investigated two deep-learning architectures: an LSTM acceptor—a network specifically tailored to encode amino acid sequences—and an autoencoder. Both models used the LSTM acceptor for peptide encoding, while the outputs from all encoders were fed into a multilayer perceptron to predict the binding probability. Building on ERGO, ERGO-II~\cite{springer_contribution_2021}  further integrated additional biological information, such as the CDR3 $\alpha$ segment, MHC typing, V and J genes, and the T cell type (CD4+ or CD8+), which led to an increase in accuracy, especially for previously unseen peptides. The authors demonstrated that the performance of ERGO-II was predominantly driven by the $\beta$ chain CDR3 sequence, followed by the V and J segments of the $\beta$ chain, and the $\alpha$ chain, in that order. The MHC allele contributed the least to the model's predictions.

Published in 2021, TITAN~\cite{weber2021a}, a supervised neural network inspired by drug sensitivity prediction models~\cite{Manica2019,born2021datadriven}, employed convolutional layers, self-attention, and multi-head context-attention layers.
Importantly, TITAN experimented with encoding peptides with SMILES (Simplified Molecular Input Line-Entry System), a linear and readable format to represent molecules atom-wise, which enables an efficient token-based input of atomic units into a neural network. Encoding epitopes as SMILES strings brought two significant benefits. First, due to the multiple paths for traversing a molecular graph, the same molecule can be encoded in various equivalent ways as a SMILES string, a property that was leveraged in TITAN as an effective data augmentation strategy. Second, it enabled the pretraining of the network using protein-compound interactions, resulting in a substantial enhancement of the model's performance. This is an example of transfer learning, where large amounts of related data are used to improve predictions in scenarios with limited data availability. 
Interestingly, while SMILES is widely used for molecule representation in machine learning, alternative embeddings like SELFIES (Self-Referencing Embedded Strings)~\cite{krenn_self-referencing_2020} offer improved performance in correctness and encoding physical constraints. SELFIES are gaining traction in generative chemistry models, where producing computationally valid molecules is essential, \deleted{. However, to our knowledge, SELFIES have not yet been applied to protein modeling or protein generative tasks,} and present an intriguing direction for future exploration.

Also based on neural networks, physics-based approaches aiming to predict the binding energy (rather than the binding probability) between a TCR and a pMHC have been proposed~\cite{xu_immunological_2018}.
This is achieved through a theoretical model that 
calculates the binding energy of a pair as the aggregate of two components: The first describes the pairwise interaction between the amino acid sequences of the CDR3 and the peptide, defined by their statistical potential in the Miyazawa-Jernigan matrix~\cite{miyazawa_residue_1996, PhysRevLett.79.765}; and the second is a constant term representing the interaction between the pMHC and the CDR1 and CDR2 regions.
Successful immunological recognition is predicted if the binding energy exceeds a certain threshold, reflecting the thymic selection process.
%
%Two binary encoding schemes were utilized for amino acid sequences: one that considers the relative interaction strengths of amino acids and another that normalizes these differences.

The key insights gained from these models are that a good latent sequence representation can significantly improve a model's
performance.
Additionally, transfer learning can further boost the model's accuracy. The pretrained PLMs developed in the 3rd generation, discussed in the next section, build upon these principles.

\subsection{3rd generation: Protein language models.} \label{sec:languageModelsImmuno}

%\noindent \emph{Transfer learning in natural language processing:}
Recently, attention-based Transformer models have garnered substantial public interest due to the release of language models fine-tuned for conversational tasks, such as ChatGPT, Bard, Perplexity, and others. Although these AI-based chatbots have brought the field into the public eye, the use of self-supervised pre-training on unlabeled data has been driving advancements in NLP for years.

Attention-based Transformers were first introduced in 2017~\cite{vaswani2017}. It soon became evident that these architectures could be trained on millions of lines of unlabeled text using proxy tasks such as masked word prediction or next word prediction~\cite{devlin2019, radford2018}. Following the pre-training phase, the models are typically fine-tuned with labeled datasets for specific text-based downstream tasks.

One of the key factors contributing to the success of these models was the transition from one-hot encodings—a simplistic transformation that converts text into vectors without preserving any semantic or contextual word information—to more sophisticated models. The first significant advancement came with Word2Vec~\cite{mikolov2013}, which took into account the local context of words to generate continuous distributed representations. Word2Vec was able to capture semantic relationships between words based on their co-occurrence within a fixed window size, providing a more nuanced representation than one-hot encoding, though it lacked dynamic context sensitivity.

More recently, Transformer architectures such as GPT (Generative pre-trained Transformer)~\cite{radford2018, radford2019, brown2020}, BERT (Bidirectional Encoder Representations from Transformers)~\cite{devlin2019}, Transformer-XL~\cite{dai_transformer-xl_2019}, and XLNet~\cite{yang_xlnet_2020} have gained prominence.
Transformer architectures, trained on extensive text corpora, encounter words in a variety of contexts, which allows them to be context-aware. By leveraging attention mechanisms, they can capture broader contextual dependencies across the entire input sequence. These models excel in learning rich contextual representations by processing sequences either bidirectionally or unidirectionally with attention mechanisms, enabling them to discern intricate relationships between words in diverse contexts.s

\noindent \emph{Learning the language of biology:}
With the success of Transformer architectures in NLP tasks, they were quickly adapted to various biological tasks, such as biomedical text mining~\cite{lee_biobert_2020} and genomic sequence analysis~\cite{ji_dnabert_2021}. In the context of protein modeling, the sequential order of amino acids in proteins follows rules determined by chemical properties, such as polarity, charge, and hydrophilicity. This is analogous to how grammatical rules govern the arrangement of words in sentences.
Indeed, since the year 2020, a variety of Transformer models have emerged that, trained on extensive protein sequence datasets, are demonstrating competitive performance in various protein-related tasks, such as predicting protein homology, structure, function, and interactions~\cite{nambiar2020,  rives2021, elnaggar_prottrans_2022, brandes2022,  chowdhury_single-sequence_2022,madani_large_2023,lin2023}.

\subsubsection{Language models for TCR specificity prediction}~\label{sec:languageModelsImmuno2}
With PLMs becoming increasingly popular, the first attempts to apply them to the TCR specificity prediction problem were soon made. For example, TCR-BERT~\cite{wu2021} is a BERT-based model that was pre-trained on a dataset of 88,403 TCR$\alpha$ and TCR$\beta$ sequences using a masked token prediction task. The weights were fine-tuned on an antigen classification task, and the resulting TCR embeddings were reduced to 50 dimensions using principal component analysis (PCA), followed by classification with a support vector machine (SVM).

More recently, STAPLER~\cite{kwee2023}, another BERT-based architecture, was trained on an even larger unlabeled dataset of almost 80M random pairs of TCR$\alpha$, TCR$\beta$, and peptide sequences using similarly a masked token prediction task.

Although not a Transformer model, CatELMo~\cite{zhang2023} exploits ELMo (Embeddings from Language Models)~\cite{peters_deep_2018}, a bi-directional context-aware LSTM-based language model, for TCR modeling. CatELMo is trained on more than four million TCR sequences collected from ImmunoSEQ~\cite{nolan2020}, and achieves improved performance compared to traditional TCR and epitope sequence embeddings, such as BLOSUM.

All these models leverage transfer learning by utilizing unlabeled TCR sequences for pre-training. Transfer learning can also be utilized by leveraging existing PLMs that have been trained on vast collections of unlabeled protein sequences, not just TCR sequences. 
For instance, TCRconv~\cite{jokinen2023}  employs ProtBERT~\cite{elnaggar_prottrans_2022} embeddings as input and processes them through a convolutional and a linear layer for the downstream task of epitope classification. 
When considering using a pretrained PLM, a relevant question is whether it is better to use a large, general PLM or a domain-specific PLM trained exclusively on TCR sequences.
Recent studies have compared general and domain-specific PLMs in the context of antibody property prediction~\cite{nijkamp_progen2_2023, wang2023on, harmalkar_toward_2023, deutschmann_domain-specific_2023}, however, the findings are not easily generalizable due to differences in the predictive tasks and models versions. For instance, while the ProGen2 models~\cite{nijkamp_progen2_2023} pretrained on universal proteins outperformed antibody-specific models,  antibody-specific PLMs excelled in high-specificity tasks compared to ESM1, a general PLM~\cite{wang2023on}. However, the comparison may now be outdated given the advancements of ESM2~\cite{lin2023} over ESM1~\cite{rives2021biological}.

\begin{figure}[ht!]
\centering
\includegraphics[width=\linewidth]{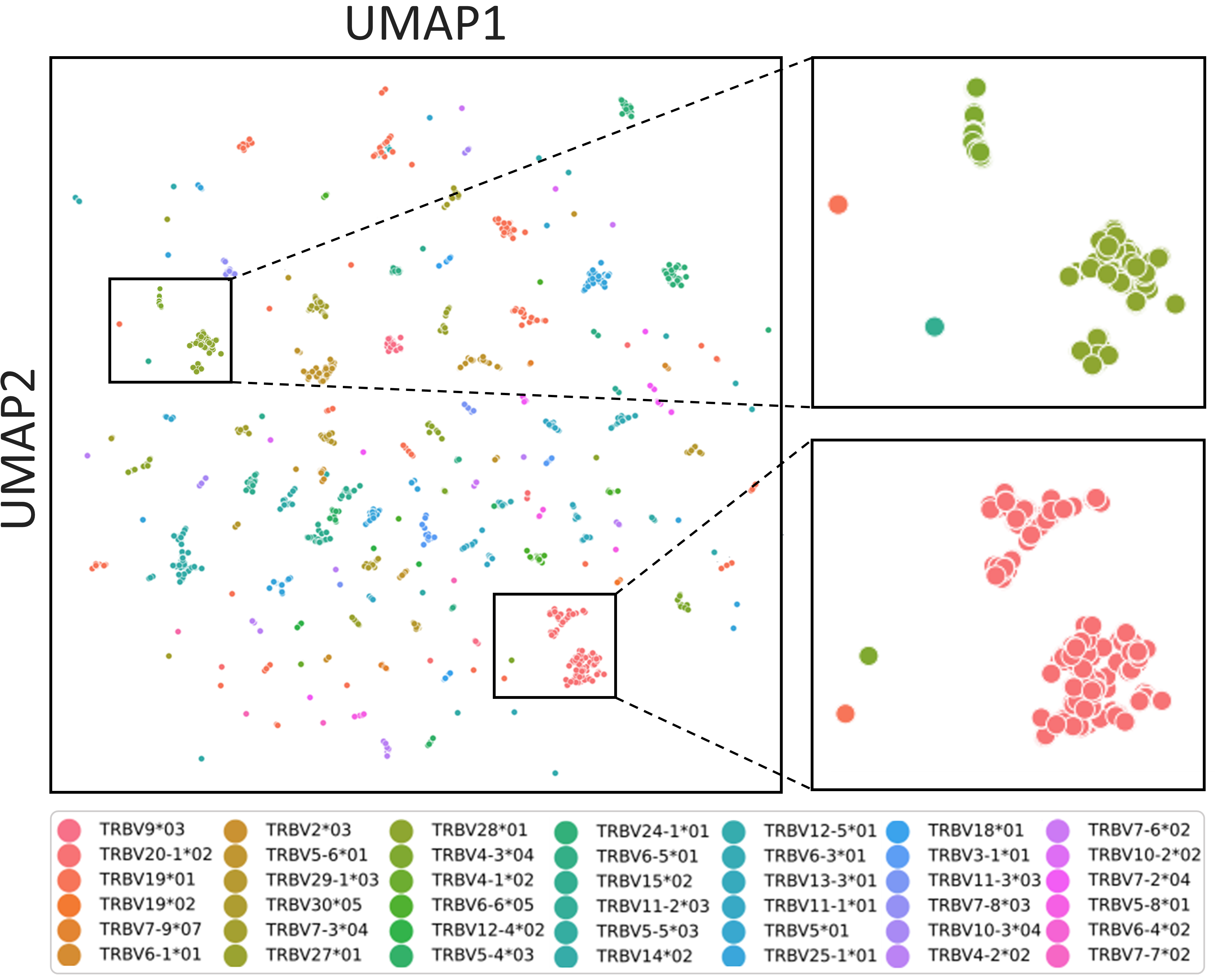}
\caption{\textbf{ESM2 embeddings of the TCR$\beta$ chain sequence}, as reported in~\cite{meysman2023}, demonstrate the effectiveness of pretrained Transformer models to capture biologically relevant structures within their latent space. Each sequence's embedded representation is projected into 2D using UMAP. In the figure, the ESM2 latent space naturally clusters sequences based on the usage of the V-segment gene. }
\label{fig:biolanguage}
\end{figure}

Further nuances were added to the debate by demonstrating that the choice between general and domain-specific PLMs depends on the task and data availability. For instance, a recent study showed that  ESM2~\cite{lin2023}  (Fig. \ref{fig:biolanguage}) and TCR-BERT~\cite{wu2021} yielded comparable results in a TCR classification task~\cite{deutschmann_domain-specific_2023}. However, the same study showed that a general model outperformed a domain-specific PLM as the amount of data available for finetuning increased. 
Furthermore, the optimal model architecture, including the number of layers and model size, can vary depending on the specific task and dataset size. The conclusion was that, as each PLM encodes information differently, the selection of PLM embeddings and the design of the coupled architecture must be carefully optimized for each downstream application.

%This underscores the need for further research to identify the best practices for TCR prediction tasks.

%A recent study showed that embeddings from ESM2~\cite{lin2023}, a general PLM (Fig. \ref{fig:biolanguage}), and TCR-BERT~\cite{wu2021} yielded comparable results in a TCR classification task~\cite{deutschmann_domain-specific_2023}.
%
%This study further showed that general models trained on broader collections of heterogeneous protein sequences can outperform their domain-specific counterparts if there is enough data to fine-tune them for the downstream task.
%This finding challenges the prevailing assumption that domain-specific models offer superior performance in specialized tasks.  Indeed, as each PLM encodes information differently, the selection of PLM embeddings and the design of the coupled architecture have to be carefully optimized for each downstream application. 

\noindent \emph{Limitations:} While powerful, Transformer models also have potential limitations. A central challenge for contextualized language models is their ability to recognize lexical ambiguity, i.e. amino acid sequences that possess multiple meanings. Two principal types of lexical ambiguity are polysemy and homonymy. Within PLMs, polysemy refers to identical amino acid sequences or motifs that perform different functional roles based on their context. For example, a motif might function as a signaling domain in one protein but serve as a structural domain in another. Homonymy, on the other hand, refers to distinct sequences that have evolved to appear identical but have different functions in different protein contexts, i.e. identical motifs in different proteins or organisms that have different functions. If a PLM cannot effectively capture the contextual nuances, it may lead to a \emph{ meaning conflation deficiency}, where the model is unable to distinguish between the different meanings of similar sequences. This can negatively impact the model's performance and accuracy in tasks like predicting protein functions, interactions, or structures. The effects of homonymy and polysemy have been explored in natural language. For instance, recent work showed that BERT could distinguish between completely different meanings (homonyms), however, it struggled to identify the nuances of polysemic words~\cite{haber_patterns_2021}. Similar studies in the context of PLMs are still needed to assess their capabilities in handling lexical ambiguity.

\subsection{Interpretability}
%The interpretation of existing models is also necessary to improve our understanding of TCR-epitope interactions.
%
Many machine learning models, especially PLMs, are inherently non-interpretable due to the encoding of amino acid information in highly abstract and complex latent spaces. Lack of understanding about a model's inner behavior in high-stakes scenarios is dangerous, as it limits trust in the model and its \emph{black-box}  nature might hide data biases and algorithmic errors. %However, many recent approaches to investigate and extract biological meaning from TCR black box models and PLMs have been recently proposed. 
Despite their robust performance, transformer models face challenges with interpretability. If unaddressed, these challenges might potentially hide model biases and spurious correlations~\cite{niven_probing_2019,kurita_measuring_2019}. The first PLMs already delved into the question of interpretation. For instance, early efforts showed that the output embeddings from a pretrained Transformer could reflect proteins' structural and functional attributes via learned linear transformations~\cite{rives2021biological}. Later on, various works analyzed output embeddings of protein models through dimensionality reduction techniques, such as PCA or t-SNE~\cite{elnaggar_prottrans_2022, biswas_low-n_2021}. 

More recently, NLP researchers have been focusing on interpreting Transformers. For instance, the emerging field of Bertology~\cite{rogers_primer_2020} aims to unravel the workings of the BERT model~\cite{devlin2019} and to understand its effectiveness and internal mechanisms. This includes dissecting the model's layers, attention mechanisms, embedding spaces, and its approach to capturing linguistic and contextual information. In this context, a novel application of Bertology to PLMs investigated the diverse biological concepts learned by different attention heads, analyzing the self-attention weights across various Transformer architectures and protein datasets~\cite{vig_bertology_2020}. The authors found that attention maps correlate with amino acid substitution matrices and can identify protein contact maps, binding sites, and post-translational modifications. Intriguingly, different attention layers identified distinct information, supporting earlier NLP findings that deeper layers in text-based Transformers focus on more complex attributes and encode sophisticated representations~\cite{vig_analyzing_2019, raganato_analysis_2018}.

Other modifications of attention layers inspired by statistical physics have been proposed. For instance, factored attention is an energy-based attention layer that merges Transformers and energy-based  Potts model, a type of Markov Random Field approach typically used for modeling interactions between amino acids. The authors introduced a simplified energy-based attention model trained on alignments, which interpolated between the standard attention mechanism and Potts models. The model provided a granular understanding of how Transformers process data, especially in capturing complex hierarchical signals in protein family databases~\cite{bhattacharya_interpreting_2021}.

Representation learning, which focuses on transforming raw data into a concise, high-level format that captures essential features, is becoming increasingly significant in biology, particularly for analyzing protein sequences. Recent work showed that taking representation geometry into account significantly improves interpretability and lets the models reveal biological information that is otherwise obscured~\cite{detlefsen_learning_2022}.

As NLP and Bertology thrive, numerous new methods for interpreting language models emerge. However, applying these methods directly to PLMs is challenging due to the fundamental differences between natural language and amino acid sequences. 
As an example, self-consistency approaches~\cite{wang_self-consistency_2022} proposed in NLP, which use chain-of-thought prompting~\cite{wei_chain--thought_2022} to explore multiple reasoning paths, might not be suitable for PLMs, which focus on physical and chemical interactions rather than logical reasoning. The complexity and high-dimensional nature of protein representations further complicate the direct application of NLP techniques to PLMs. Nonetheless, adaptations might be possible, such as using models to explore protein conformations or interactions.
Other approaches such as Automated Concept-based Explanation (ACE)~\cite{ghorbani_towards_2019}, 
which aims to identify key concepts, i.e. \emph{high-level, human-understandable features} influencing model decisions, could be tailored for PLMs to elucidate the biochemical properties governing protein behaviors, offering a bridge to more interpretable and insightful analyses in the realm of protein functions.

Beyond interpreting PLMs, attention mechanisms can also be leveraged to capture structurally important
residue pairs that contribute to TCR-epitope binding~\cite{kim_tspred_2023, dens_interpretable_2023} or predict protein structural properties~\cite{koyama_attention_2023}. For instance, a recent study categorized residues into groups with high and low attention values. The analysis revealed that residues receiving more attention often had distinct structural properties, and they were statistically more likely to form hydrogen bonds within the CDR3 region~\cite{koyama_attention_2023}.
%
%Alternatively, black-box models can be employed to generate plausible hypotheses that are later investigated using post hoc interpretable methods. An example of this approach is DECODE~\cite{papadopoulou2022},  a user-centric interpretable pipeline to extract human-comprehensible rules from any black-box TCR predictive model. DECODE exploits Anchors, a model-agnostic approach to approximate the decision boundary of any machine learning model and identify local, sufficient conditions for predictions~\cite{ribeiro_anchors_2018}.

Finally, it is also worth mentioning post hoc methods that aim to help understand the predictions of a black-box model after they have been made. Popular methods include LIME~\cite{ribeiro_why_2016}, Anchors~\cite{ribeiro_anchors_2018}, or SHAP~\cite{lundberg_unified_2017}. Focusing on TCR classifiers, specific approaches have been developed for the post hoc interpretation of black-box TCR predictive models. For instance, DECODE~\cite{papadopoulou2022} is a user-centric interpretable pipeline designed to extract human-comprehensible rules about a model's predictions. It utilizes Anchors~\cite{ribeiro_anchors_2018}, a model-agnostic method to approximate the decision boundary of any machine learning model and identify local, sufficient conditions for predictions.

A caveat of post-hoc approaches is that they produce \emph{local} explanations. This means they cannot explain the model as a whole but rather provide explanations for predictions about individual instances or data points.
It has been proposed that transparent models should be preferred~\cite{rudin_stop_2019}, however, the discussion between the need for transparency and the accuracy trade-off when handling complex abstract data remains open. 
As a middle ground, architecturally constrained neural networks, which process each input feature independently in a manner akin to linear models, have been suggested to improve the interpretability of the TCR binding prediction task~\cite{nguyen_flan_2023}. 

In summary, the field of interpretable deep learning is rapidly evolving, and numerous models and approaches are continually emerging. While more research is needed to identify which methods are suitable for understanding TCR black-box models and PLMs-based models, 
interpreting model predictions is not just a technical necessity but a fundamental requirement for clinical trust and application~\cite{gilpin2019, linardatos2021, dens2023interpretable}.

\section{Outlook}

The accurate prediction of TCR specificity is crucial for various clinical applications, including the development of safer and more efficient immunotherapies and gaining a deeper understanding of autoimmune diseases. 
Despite the progress, current TCR prediction models face significant challenges due to insufficient and often biased data, notably concerning epitope information. This limitation restricts the models' capacity to generalize to new epitopes, which is critical, for instance, to predict potential cross-reactive events in T~cell-based therapies.

In recent years, the field of immunoinformatics has significantly benefited from the integration of machine learning techniques with proven efficacy in other domains, including established neural network architectures such as Convolutional Neural Networks (CNNs) and Recurrent Neural Networks (RNNs), transfer learning strategies, and NLP-inspired approaches such as word embeddings. The recent adoption of Transformers and language models has led to the emergence of protein language models (PLMs), which have substantially improved the accuracy of computational models for protein analysis and prediction, including TCR prediction models. This innovation is revolutionizing performance in data-scarce environments and demonstrating remarkable successes across various protein-centric tasks. Despite these advancements, considerable work remains to be done, particularly in enabling these models to reliably execute highly specific functions, such as predicting TCR specificity to novel epitopes.

The IMMREP22 benchmark~\cite{meysman2023} represented an initial effort to provide a quantitative evaluation of various  TCR prediction models. However timely and much-needed, the benchmark also underscored the difficulties in making fair comparisons across machine learning models. 
For instance, challenges such as the lack of ground truth for negative TCR-epitope data~\cite{dens2023pitfalls}, the inclusion of similar TCR sequences, limited dataset sizes, the necessity to train models using the same data—penalizing models that exploited transfer learning or pretraining strategies—and the use of the same data for both training and evaluation, which limited the testing of model generalizability, added layers of complexity to comparisons across models.
Moreover, robust comparisons across methods are complicated by the different methodological designs, with some approaches using only the CDR3 $\beta$ chain, which has more data available, while others use the full paired $\alpha$ and $\beta$ chains, offering potentially better predictions~\cite{meysman2023} but less training data. Additionally, some methods train on both paired and unpaired data, allowing for predictions in both scenarios~\cite{fast_tapir_2023}.
Due to these challenges, the IMMREP22 benchmark report~\cite{meysman2023} emphasized overarching performance trends rather than the results of individual models. \added{A more recent TCR benchmark, IMMREP23, organized on Kaggle~\cite{immrep23_kaggle}, overcame most of the prior challenges but faced new unexpected ones, including a partial release of the test data during the competition and a data leakage identified and exploited by some participating teams. Rather than invalidating the utility and value of these competitions, this highlights the difficulty of constructing truly unbiased benchmarks. }
%A more recent TCR benchmark, IMMREP23 organized on Kaggle~\cite{immrep23_kaggle}, has demonstrated the effectiveness of Transformer-based models in TCR prediction. 
While waiting for the report to be published, in this review we have chosen to discuss methodologies, general trends, and biases following the IMMREP22 approach, rather than discussing individual model performances.

Finally, many computational methods designed for TCR specificity prediction can be adapted to model B cell receptor (BCR) antigen binding. Indeed, both statistical~\cite{marcou_high-throughput_2018}, machine learning~\cite{isacchini_deep_2021}, and Transformer-based approaches~\cite{zhao_sc-air-bert_2023} have been developed to jointly model TCR and BCR repertoires. While not the primary focus of this paper, BCR specificity prediction faces similar data and computational challenges as TCR specificity prediction. However, BCR-antigen binding typically involves non-linear epitopes, unlike the linear epitopes involved in T-cell peptide binding. Furthermore, BCR datasets often list the target antigen with an unknown epitope, making the BCR problem generally more challenging. Addressing these challenges might involve adapting TCR models for BCR modeling by modifying the algorithms to incorporate structural information about non-linear epitopes. Additionally, improving the resolution of antigen mapping, employing advanced algorithms capable of handling full antigen sequences, and leveraging transfer learning to fine-tune models for BCR-related tasks will also be beneficial. Ultimately, an intriguing question remains whether PLM-based models can effectively capture both TCR and BCR characteristics or if specialized models might yield higher accuracy.
Similarly, the question of how PLMs organize information in the latent space requires deeper investigation. A better understanding of what and how information is encoded could lead to more focused and efficient development and deployment of models for both TCR and BCR-related tasks.

In summary, the field of computational immunology, particularly in TCR  specificity prediction, is rapidly evolving, yet key questions and challenges persist. Addressing these issues is crucial not only for deepening our understanding of the immune system but also for paving the way for groundbreaking clinical applications.

\section{Funding and declarations} 

The authors acknowledge funding from the European Union's Horizon 2020 research and innovation program under the Marie Skłodowska-Curie Actions (813545) and the ICT-2018-2 Program (826121). Additional support was received from the Swiss National Science Foundation through the Sinergia program (CRSII5 193832) and Project Funding (192128).

During the preparation of this work, the authors used ChatGPT to improve language clarity and readability. After using this service, the authors reviewed and edited the content as needed and take full responsibility for the content of the publication.

\section{Appendix}\label{sec:intro_appendix}

\noindent\textbf{Survey of TCR specificity prediction models (published by December 2023)}.

\begin{longtable}[h]{l|l|l|l}
    &  &          & Epitope   \\
Model & Year & Type  &  explicitly   \\
        &  &          &   encoded      \\
\hline
TCRdist~\cite{dash2017}   & 2017 & Unsupervised & No  \\
GLIPH~\cite{glanville2017}    & 2017 & Unsupervised & No  \\
DeNeuter~\cite{deneuter2018}  & 2018 & Unsupervised & No  \\
NetTCR~\cite{jurtz_nettcr_2018} & 2018 & Supervised   & Yes \\
Xu et al~\cite{xu_immunological_2018} & 2018 & Supervised   & Yes \\
TCRex~\cite{gielis2019}    & 2019 & Unsupervised & No  \\
SETE~\cite{tong_sete_2020}     & 2020 & Supervised   & No  \\
TcellMatch~\cite{fischer_predicting_2020}  &  2020 & Supervised   & Yes$^1$      \\
ImRex~\cite{moris2020current}   & 2020 & Supervised   & Yes  \\
iSMART~\cite{zhang2020}  & 2020 & Unsupervised & No  \\
ERGO~\cite{springer_prediction_2020}    & 2020 & Supervised   & Yes  \\
TCRdist3~\cite{mayer-blackwell2021a}   & 2021 & Unsupervised & No  \\
TCRGP~\cite{jokinen_predicting_2021}    & 2021 & Supervised   & No  \\
ERGO-II~\cite{springer_contribution_2021}    & 2021 & Supervised   & Yes  \\
TCRMatch~\cite{chronister2020tcrmatch}  & 2021 & Unsupervised & No  \\
pMTnet~\cite{lu2021}   & 2021 & Supervised   & Yes  \\
TITAN~\cite{weber2021a}    & 2021 & Supervised   & Yes  \\
NetTCR-2.0~\cite{montemurro2021}    & 2021 & Supervised   & Yes  \\
TCRAI~\cite{zhang_framework_2021}     & 2021 & Supervised   & No  \\
DeepTCR~\cite{sidhom2021deeptcr} & 2021 & Supervised   & No  \\
TCR-BERT~\cite{wu2021} & 2021 & Language     & No  \\
GIANA~\cite{zhang2021}   & 2021 & Unsupervised & No  \\
ELATE~\cite{dvorkin2021}      & 2021 & Unsupervised & No  \\
DLpTCR~\cite{xu2021}  & 2021 & Supervised   & Yes  \\
SwarmTCR~\cite{ehrlich_swarmtcr_2021} & 2021 & Supervised   & No \\ 
ClusTCR~\cite{isacchini_deep_2021} & 2021 & Unsupervised  & No \\
SONIA~\cite{valkiers_clustcr_2021} & 2021 & Supervised   & No \\
ATM-TCR~\cite{cai_atm-tcr_2022}  & 2022 & Supervised   & Yes  \\
TCRconv~\cite{jokinen2023} & 2022 & Language     & No  \\
PiTE~\cite{zhang_pite_2022} & 2022 & Language   & Yes  \\ 
Bi et al~\cite{bi_attention_2022} & 2022 & Supervised   & Yes \\ 
diffRBM~\cite{bravi2022}  & 2023 & Supervised   & No$^2$  \\
PanPep~\cite{gao2023}   & 2023 & Supervised   & Yes  \\
catELMo~\cite{zhang2023}   & 2023 & Language     & Yes  \\
STAPLER~\cite{kwee2023}   & 2023 & Language     & Yes \\
%copepodTCR~\cite{kovaleva_copepodtcr_2023}   & 2023 & Supervised     & No  \\
NetTCR-2.2~\cite{jensen_nettcr_2023}   & 2023 & Supervised     & Yes  \\
TCR-H~\cite{rajitha_tcr-h_2023}   & 2023 & Supervised     & Yes   \\
epiTCR~\cite{pham_epitcr_2023}   & 2023 & Supervised     & Yes \\
GGNpTCR~\cite{zhao_ggnptcr_2023}   & 2023 & Supervised     & Yes  \\ 
Rehman Khan~\cite{khan_determining_2023}   & 2023 & Language     & No  \\ 
TAPIR~\cite{fast_tapir_2023}   & 2023 & Supervised     & Yes \\ 
BERTrand~\cite{myronov_bertrandpeptidetcr_2023}   & 2023 &   Language   & Yes \\ 
MITNet~\cite{darmawan_mitnet_2023}   & 2023 &   Supervised   & No \\ 
SC-AIR-BERT~\cite{zhao_sc-air-bert_2023}   & 2023 &   Language   & No \\ 
MixTCRpred~\cite{croce_deep_2023}   & 2023 &   Language   & Yes$^3$  \\
TSPred~\cite{kim_tspred_2023}   & 2023 &   Supervised   & Yes \\
Koyama et al~\cite{koyama_attention_2023}   & 2023 &   Language   & Yes \\
TCRen~\cite{karnaukhov_tcren_2023} & 2023 & Supervised   & Yes \\ 
TEINet~\cite{jiang_teinet_2023} & 2023 & Supervised   & Yes \\ 
\added{MIX-TPI~\cite{yang_mix-tpi_2023}} & \added{2023} & \added{Supervised}   & \added{Yes} \\
\added{AVIB~\cite{grazioli_attentive_2023}} & \added{2023} & \added{Supervised}   & \added{Yes} \\
EPIC-TRACE~\cite{korpela_epic-trace_2023}   & 2023 &   Language   & Yes \\
Deutschmann et al~\cite{deutschmann_domain-specific_2023}   & 2023 &   Language   & Yes 
\label{table:models}
\end{longtable}

\noindent $(^1)$ Both an epitope classification model and a model that encodes the epitope sequences are presented.

\noindent $(^2)$ diffRBM trains two different models: one for immunogenicity prediction, which explicitly models the epitope sequence, and a second for TCR epitope specificity prediction, which does not consider the epitope sequence.

\noindent $(^3)$ The basic model is epitope-specific; however, a pan-epitope model that encodes the sequence of the epitope is also presented.

\bibliographystyle{elsarticle-num} % Or another suitable style
\bibliography{bibliography} % Replace 'yourbibfile' with the name of your BibTeX file

\end{document}